\documentclass[twocolumn,showpacs,preprintnumbers,amsmath,amssymb,prb,superscriptaddress,aps]{revtex4}

\usepackage{epsfig}
\usepackage{graphicx}
\usepackage{dcolumn}
\usepackage{bm}
\usepackage{epstopdf}
\usepackage{amsmath}
\usepackage{xcolor}
\usepackage{multirow}

\begin{document}

\title{Evidence of Lattice Strain as a Precursor to Superconductivity in BaPb$_{0.75}$Bi$_{0.25}$O$_3$}
\author{M Bharath}
\affiliation{School of Physical Sciences, Indian Institute of Technology Mandi, Kamand, Himachal Pradesh-175005, India}
\author{Jaskirat Brar}
\affiliation{School of Physical Sciences, Indian Institute of Technology Mandi, Kamand, Himachal Pradesh-175005, India}
\author{Himanshu Pant}
\affiliation{School of Physical Sciences, Indian Institute of Technology Mandi, Kamand, Himachal Pradesh-175005, India}
\author{Asif Ali}
\affiliation{Department of Physics, Indian Institute of Science Education and Research, Bhopal, Madhya Pradesh-462066, India}
\author{Sakshi Bansal}
\affiliation{Department of Physics, Indian Institute of Science Education and Research, Bhopal, Madhya Pradesh-462066, India}
\author{Ravi Shankar Singh}
\affiliation{Department of Physics, Indian Institute of Science Education and Research, Bhopal, Madhya Pradesh-462066, India}
\author{R Bindu}
\affiliation{School of Physical Sciences, Indian Institute of Technology Mandi, Kamand, Himachal Pradesh-175005, India}

\begin{abstract}
    In this work, we have investigated the precursor effects to superconductivity in BaPb$_{0.75}$Bi$_{0.25}$O$_3$ using temperature dependent
    resistivity, x-ray diffraction technique and photoemission spectroscopy. The present compound exhibits superconductivity around 11 K
    ($T_C$). The synthesis procedure adopted is much simpler as compared to the procedure available in the literature. In the temperature range
    (10\,K--25\,K) i.e. above $T_C$, our results show an increase in both the orthorhombic and tetragonal strain. The well screened features
    observed in Bi and Pb 4$f_{7/2}$ core levels are indicative of the metallic nature of the sample. The compound exhibits finite intensity at
    the Fermi level at 300\,K and this intensity decreases with decrease in temperature and develops into a pseudogap; the energy
    dependence of the spectral density of states suggests disordered metallic state. Furthermore, our band structure
    calculations reveal that the structural transition upon Pb doping results in the closing of the band gap at the Fermi level.
\end{abstract}

\keywords{bismuthates, lattice distortion, photo electron spectroscopy, superconductivity, doping}

\pacs{61.05.C-,74.25.Jb}

\maketitle

\section{\label{sec:intro}Introduction}

The lead doped BaBiO$_3${}, BaPb$_{1-x}$Bi$_{x}$O$_3${} (BPBO), is a well-known 3-dimensional perovskite, which exhibits superconductivity for
bismuth compositions in the range $0.15 \leq x \leq 0.35$\cite{climent2011}. This family of compounds has attracted much attention because of
its relatively high transition temperature ($T_C \sim $13\,K, at $x = 0.25$\cite{sleight1993}), despite the low carrier
concentration\cite{thanh1980}. Unlike the cuprates or lanthanides, BPBO family of superconductors are (i) three-dimensional, (ii) without
transition metal ions like Cu, La, etc, and (iii) non-magnetic. In the superconducting compositions, the compound is diphasic, with the
coexistence of the tetragonal and orthorhombic phases\cite{marx1992, climent2011}; the volume fraction of tetragonal phase dictates the $T_C$,
thus being the superconducting phase. Recently, Nicoletti et al.\cite{nicoletti2017} have shown that a short range CDW order exists adjacent to
the superconducting phase (\textit{i.e.}, above $\sim$ 7\,K) in superconducting BaPb$_{0.78}$Bi$_{0.22}$O$_3$. Further, recent study by Parra et
al.\cite{parra2021}, shows that an electronic reorganization into a 2D granular superconductor takes place when BPBO approaches the
superconductor-insulator transition before ultimately transforming to an insulator. Superconductivity is generally
seen adjacent to exotic symmetry-breaking ground states, like AFM and CDW orders\cite{nicoletti2017}. Usually, a complex but subtle interplay
between spin, charge, crystal and electronic structures shape the resulting electronic properties. BPBO, which manifests such emergent electronic properties, is an
ideal candidate to engineer novel superconducting devices such as quantum computing circuits\cite{kim2019}.

The crystal structure of superconducting compositions have been a major point of confusion -- Cox and Sleight\cite{cox1976} have reported that
the compound is tetragonal in the range $0.05 < x < 0.35$; Khan et al.\cite{khan1977} observe that the compound has a distorted orthorhombic
crystal structure belonging to the  $Cmm2$ phase for all compositions, whereas, Oda et al.\cite{oda1985} concluded that the compound is
orthorhombic for all compositions with $x < 0.9$ along with a structural transition from an orthorhombic to a monoclinic phase at about 160\,K
in case of BaPb$_{0.75}$Bi$_{0.25}$O$_3$; in a later study, Asano et al.\cite{asano1988} and Oda et al.\cite{oda1986} proposed that the
differences in the sample preparation techniques could result in the compounds stabilising in either orthorhombic or tetragonal phase, each
having a distinct $T_C$.

Investigations of the phase separation of BaPb$_{1-x}$Bi$_{x}$O$_3${} by Giraldo-Gallo et al. using high quality single
crystals\cite{giraldo2015} have lead to the observation that the structural dimorphism takes the form of partially disordered nanoscale stripes
for optimal composition of $x \approx 0.24$. As the bismuth composition deviates from its optimal value, the volume of tetragonal phase reduces
and the tetragonal stripes reduce to islands embedded in a matrix of orthorhombic BaPb$_{1-x}$Bi$_{x}$O$_3${}. Point-contact spectroscopic
measurements performed on BaPb$_{1-x}$Bi$_{x}$O$_3${} ($0 \leq x \leq 0.28$) indicate that the metal-insulator transition (MIT) is driven by
disorder\cite{luna2014}. Magnetoresistance studies performed on BaPb$_{0.75}$Bi$_{0.25}$O$_3$ epitaxial thin films grown on LaLuO$_3$ by Harris
et al. \cite{harris2018}, show superconducting fluctuations in epitaxial thin films well above $T_C$; and in case of the thickest films with
thickness above 100\,nm these fluctuations were observed till 27\,K.

Several photoemission spectroscopy measurement studies on BaPb$_{1-x}$Bi$_{x}$O$_3${} have been reported, some performed on single crystals
\cite{wertheim1982, matsuyama1989, winiarski1991, namatame1993}, and few more of polycrystalline samples\cite{sakamoto1987, kostikova2001,
korolkov2002}. Most studies do not reveal any charge disproportionation in the compound\cite{wertheim1982, sakamoto1987, matsuyama1989,
winiarski1991, namatame1993}, notable exception being Kostikova et al\cite{kostikova2001}. and Korolkov et al.\cite{korolkov2002}, who noticed
the presence of both Pb$^{\rm II}$ and Pb$^{\rm IV}$, along with Bi$^{\rm III}$ and Bi$^{\rm V}$ in their compound under study --
BaPb$_{0.8}$Bi$_{0.2}$O$_3$. Room temperature (RT) x-ray photoelectron spectroscopic (XPS) studies by Wertheim et al.\cite{wertheim1982} and
Sakamoto et al.\cite{sakamoto1987} show no indication of any density of states (DOS) at Fermi level ($E_F$). However, later ultraviolet
photoemission studies (UPS) by Matsuyama et al.\cite{matsuyama1989} (at RT and 180\,K), and Namatame et al.\cite{namatame1993} (at
liquid-nitrogen temperature (LNT)) indicate a finite DOS at $E_F$. The UPS studies by Matsuyama et al.\cite{matsuyama1989} show that
BaPb$_{0.85}$Bi$_{0.15}$O$_3$ has a clear Fermi-edge structure characteristic of normal metal; suggesting that the superconductivity would be
driven by Cooper pairing of electrons in the Fermi-liquid states. According to the BCS theory, good metals do not become superconductors because
of weak-electron phonon coupling, which makes their claim a very interesting one. It is also important to note that their measurements were done
at 180\,K, a temperature significantly higher than the $T_C$ of the compound. The electronic structure studies by Namatame et
al.\cite{namatame1993} was performed at LNT in an attempt to obtain a picture of the electronic structure of BaPb$_{1-x}$Bi$_{x}$O$_3${} for a
wide range of compositions. Further, studies exploring effects such as superconducting fluctuations discovered by Harris et al.\cite{harris2018}
indicate a presence of precursor effects before the onset of bulk superconductivity. To understand the manifestation of co-existence of two
structural phases in the electronic structure or precursor effects to superconductivity in BaPb$_{0.75}$Bi$_{0.25}$O$_3$ we have carried out
temperature dependent crystal structure and electronic structure studies on the compounds above $T_C$.

Another interesting aspect of the BaPb$_{1-x}$Bi$_{x}$O$_{3-\delta}${} series is the strong dependence of the superconductivity on the oxygen
content in the sample\cite{hashimoto1993}. The authors found that an oxygen deficiency of greater than $\delta = 0.11$ destroyed the
superconductivity. Hence, an elaborate preparation procedure is generally used to ensure that sample is well-oxygenated. The typical preparation
route is via solid state reactions of powdered raw materials -- BaCO$_3$, Pb$_3$O$_4$ and Bi$_2$O$_3$ -- taken in stoichiometric proportions.
These raw materials are mixed in ethanol and ground well using a ball-milling machine. The mixture is then heated at 720 $^\circ \rm{C}$ for
12\,h under flowing oxygen. Further, the calcined sample is ground again, pressed into a pellet and then sintered at 800 $^\circ \rm{C}$ for
12 h, in oxygen atmosphere.

In this paper, we investigate the possible precursor effects leading to the superconductivity. We present a detailed investigation of
temperature dependent resistivity, crystal structure and electronic structure studies on the compound under study. The electrical transport and
magnetic susceptibility measurements indicate a superconducting transition around ~11\,K. Further, these studies reveal that the disorder plays
a very important role in the electrical resistivity and the electronic structure of the compound. The spectral DOS near the Fermi level displays
a square-root of energy dependence indicating the compound to be a disordered metal above $T_C$. The temperature evolution of the symmetrised
spectral density of states indicates the development of a disorder-induced pseudogap at the Fermi level. Band structure calculations performed
using the TB-mBJ correlation functional indicates that the monoclinic $I2/m$ to tetragonal $I4/mcm$ structural phase transition upon Pb doping
of the parent compound plays a major role in the closing of the band gap observed in the compound under study. Additionally, we have
adopted a much simpler synthesis procedure as compared to the ones available in the literature.

\section{\label{sec:exptl}Experimental}

The polycrystalline samples of BaPb$_{0.75}$Bi$_{0.25}$O$_3${} were prepared via solid state route from BaCO$_3$, Bi$_2$O$_3$ and Pb$_3$O$_4$.
The raw materials were first preheated at 450 $^\circ \rm{C}$ for 2 hours to remove any traces of moisture absorbed by them. The proper
stoichiometric ratios of the raw materials were then ground thoroughly in a agate mortar for 24-48 hours. This finely ground powder was then
calcined in a box furnace at 900\,$^\circ \rm{C}$ for 48 hours. The heating rate was maintained at 5\,$^\circ \rm{C}$/min and the samples were
allowed to cool down to room temperature naturally inside the box furnace. The calcined powders were then sintered in the box furnace at
900\,$^\circ \rm{C}$ for around 150 hours, with intermediate grindings every 24 hours. The final sintering was performed at 950\,$^\circ
\rm{C}$. After each grinding, the powders were then pressed into pellets of 10 mm diameter under a hydrostatic pressure of 5 tons. It was found
that covering the pellets with a thin layer of powder of the sample ensured a good stoichiometric compound, and prevented the escape of Bi or Pb
from the surface of the pellet. It is to be noted that despite not sintering the sample in oxygen atmosphere, we were able to obtain a
superconducting transition.


The temperature dependent \textit{xrd} (T-XRD) measurements were performed using Rigaku's Smart Lab x-ray diffractometer powered by a 9kW
rotating anode x-ray generator. The T-XRD patterns were collected using Cu K$_\alpha$ radiations and were collected in the 2$\theta$ range of
19$^\circ$ to 87$^\circ$ at a scanning speed of 1$^\circ$ per minute with a step size of 0.02$^\circ$. The Magnetic Properties Measurement
System (MPMS) from Quantum Design, Inc. was used in the measurement of temperature dependent dc magnetization measurements. The measurement was
performed at an applied magnetic field of 0.5 T, in the temperature range of 300K to 2K. Electrical resistivity measurements were performed on
the sample using Physical Properties Measurement System (PPMS) setup from Quantum Design, Inc, in the same temperature range, using a standard
4-probe DC resistivity method. The probes were attached to the sample using high quality silver paste, and the data collection was done during
the cooling cycle. The field emission scanning electron microscopy (FE-SEM) micrographs were procured using Nova Nano SEM-450 at room
temperature at a chamber pressure of $10^{-5}$\,mbar, with a scanning voltage of 10\,kV.

The temperature dependent photoemission spectra were collected on Scienta R4000 hemispherical analyser using a monochromatic Al K$_\alpha$
(1486.6 eV) x-ray source, He I (21.2 eV), and He II (40.8 eV) ultraviolet radiations at various temperatures between 300\,K to 30\,K. The
binding energy scale was calibrated by measuring the Fermi level of Ag pellet, cleaned \text{in situ} by argon ion sputtering, using
monochromatic Al K$_\alpha$ and He$_{\mathrm I}$. A clean sample surface was obtained by fracturing the mounted samples in a chamber having
vacuum better than $7 \times 10^{-11}$\,mbar. The base pressure during the measurement was $5 \times 10^{-11}$\,mbar.

\section{\label{sec:comptl}Computational}

Band structure calculation using self-consistent full potential linear augmented-plane-wave (LAPW) were performed for
BaPb$_{0.75}$Bi$_{0.25}$O$_3${} using the code implemented in Elk\cite{elk}. We have used local density approximation (LDA)\cite{perdew} for the
exchange potential and Tran-Blaha modified Becke-Johnson (TB-mBJ) potential\cite{tran} for the correlation part. The calculations were performed
using the structural parameters obtained from the refinement of the \textit{xrd}{} patterns of $I4/mcm$ phase of BaPb$_{0.75}$Bi$_{0.25}$O$_3${}
taken at room temperature. A $1\times1\times1$ supercell was constructed and one atom of Pb was replaced by Bi, to account for the 75\% doping
of Pb at the Bi-site. The muffin-tin shape approximation for the potential well in the crystal lattice employed in the calculations used radii
of 2.8, 2.56, 2.43 and 1.43 bohr for Ba, Bi, Pb and O, respectively. The difference in total energy required for the termination of
self-consistent cycles was set to be less than $10^{-4}$ Hartree/cell. In order to understand the evolution of the states at $E_F$, we have also
performed TB-mBJ calculations on BaBiO$_3$ using the structural parameters of BaPb$_{0.75}$Bi$_{0.25}$O$_3${}, using a muffin-tin radii of 2.8,
2.29 and 1.72 bohr for Ba, Bi and O, respectively.

\section{\label{sec:results}Results and Discussions}

\subsection{\label{subsec:genchar}General Characterization}


\begin{figure}
    \centering
    \includegraphics[width=\columnwidth]{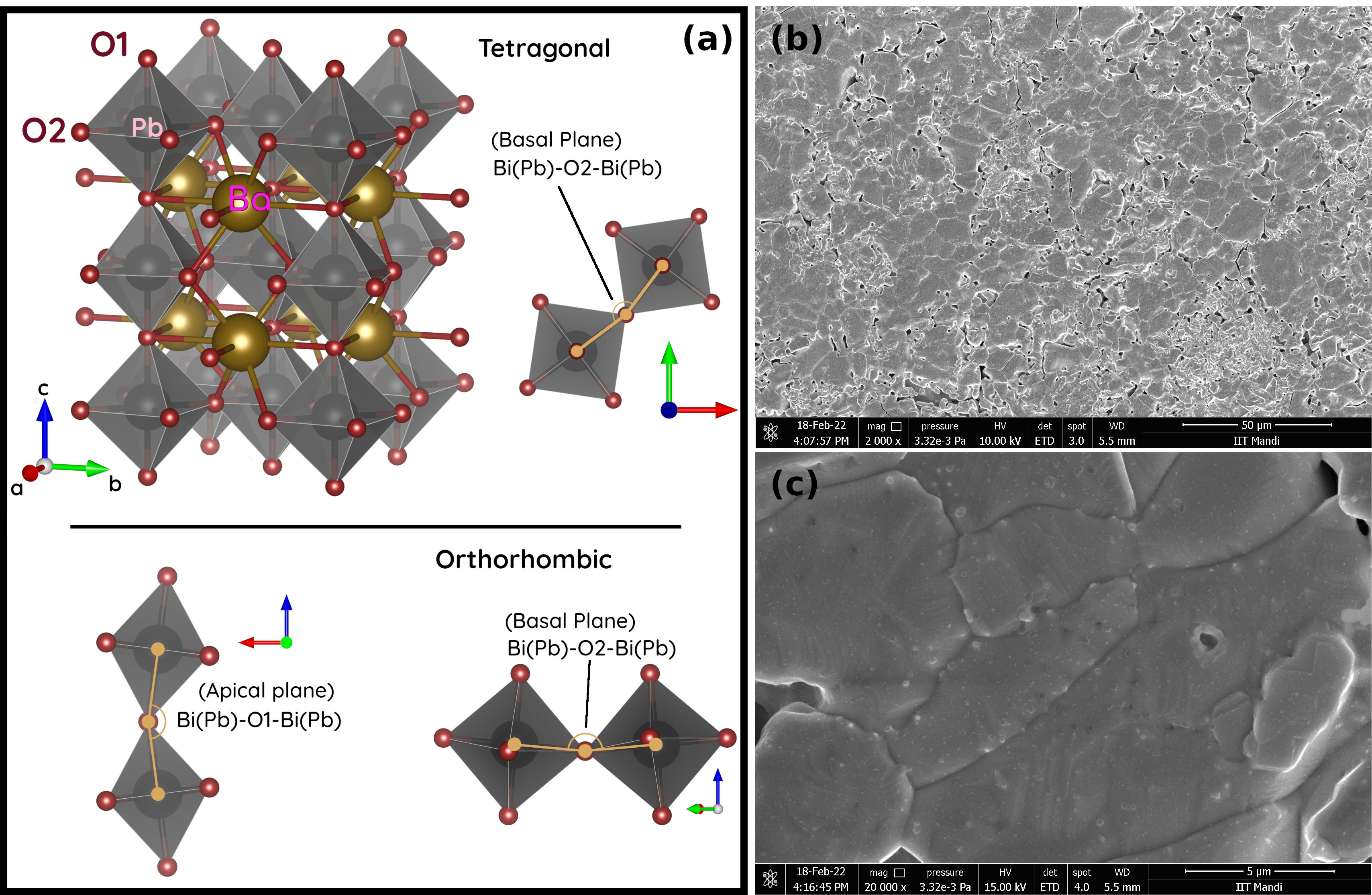}
    \caption{
        (Color online)
        (a) Crystal structure of BaPb$_{0.75}$Bi$_{0.25}$O$_3$ -- upper panel shows the structure of the tetragonal phase
	    along with the Bi(Pb)-O$_2$-Bi(Pb) angle between the two Bi(Pb)-O$_6${} octahedra in the basal plane. The lower panel shows the
	    Bi(Pb)-O$_1$-Bi(Pb) angle along the apical plane and Bi(Pb)-O2-Bi(Pb) angle along the basal plane in case of orthorhombic phase.
        (b), (c) show the FE-SEM micrographs of BaPb$_{0.75}$Bi$_{0.25}$O$_3$ sintered pellets taken at 2000x and 20000x magnifications,
	    respectively.
    }
    \label{fig:bpbo-str-sem}
\end{figure}

Figure~\ref{fig:bpbo-str-sem}(a) shows the crystal structure of the compound under study; also shown are the Bi(Pb)-O-Bi(Pb) angles of the
apical and basal planes of the two phases. Panels (b) and (c) for the same figure (\ref{fig:bpbo-str-sem}) show FE-SEM micrographs of the
BaPb$_{0.75}$Bi$_{0.25}$O$_3$ pellet. The average grain size was estimated to be $\sim$8\,$\mu$m.

\begin{figure}
    \centering
    \includegraphics[width=0.80\columnwidth]{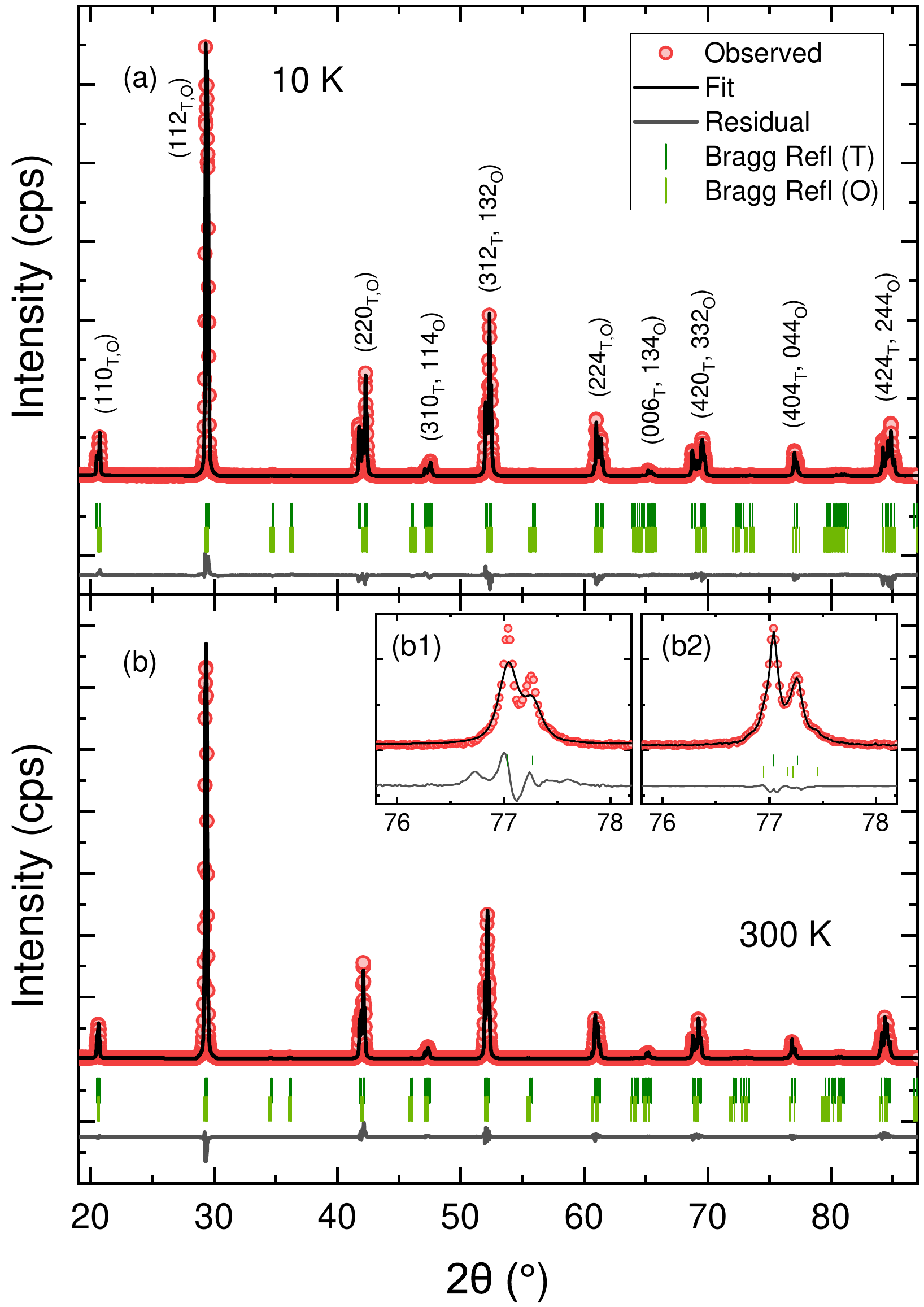}
    \caption{
        (Color online) Rietveld refinement of \textit{xrd}{} patterns collected at (a) 10\,K and (b) 300\,K. The fitting is performed using two
        phases -- I4/mcm and Ibmm -- at both temperatures. The dark green tics indicate the Bragg reflections arising from the tetragonal I4/mcm
        phase, and the light green tics from the orthorhombic Ibmm phase. Inset (b1) and (b2) show the fit of the peaks in the range $76^\circ
        \leq 2\theta \leq 78^\circ$ when modelled using a single I4/mcm phase, and two phases (I4/mcm and Ibmm) respectively.
    }
    \label{fig:bpbo-txrd-rv}
\end{figure}

In figure~\ref{fig:bpbo-txrd-rv}, we show the typical Rietveld fitting of the \textit{xrd}{} patterns of BaPb$_{0.75}$Bi$_{0.25}$O$_3$ collected
at 10\,K and 300\,K. A two-phase Rietveld refinement, consisting of tetragonal I4/mcm phase and orthorhombic Ibmm phase, has been performed to
fit the experimental data; the crystal structure and the room temperature lattice parameters are in line with the earlier works\cite{marx1992,
climent2011, giraldo2015, nicoletti2017}. The absence of unindexed peaks points to the purity of the sample. The lattice parameters and atomic
positions obtained from the refinement of the patterns collected at 300\,K and 10\,K are shown in Table~\ref{tbl:xrd-rv}.

\begin{table}[hbt]
    \caption{
        The lattice parameters, and atomic positions of I4/mcm and Ibmm phases obtained from the two-phase refinement of the spectra collected
        at 300\,K and 10\,K. The Wyckoff positions of Ba is 4b, Bi/Pb is 4c, O1 is 4a and O2 is 8h in tetragonal phase. In case of the
        orthorhombic phase these atoms occupy 4e, 4a, 4e, and 8g positions, respectively.
    }
    \label{tbl:xrd-rv}

    \begin{tabular}{ccccccc} \hline\hline
        $\qquad$   &   $\qquad$    &   \multicolumn{2}{c}{Tetragonal Phase}  &   $\qquad$    &   \multicolumn{2}{c}{Orthorhombic phase}      \\
                   &               &   10\,K        &           300\,K       &               &   10\,K       &     300\,K     \\ \cline{3-7}
                                                                                                                              \\
        \multicolumn{7}{l}{Lattice parameters}                                                                                \\
        a (\textup{\AA})  &        &   6.02815      &     6.05055            &               &   6.07851     &    6.08959     \\
        b (\textup{\AA})  &        &   -            &     -                  &               &   6.04651     &    6.06051     \\
        c (\textup{\AA})  &        &   8.62396      &     8.60776            &               &   8.51402     &    8.55560     \\
                                                                                                                              \\
        \multicolumn{7}{l}{Atomic Positions}                                                                                  \\
        \multicolumn{1}{l}{Ba}  &   x     &     \multicolumn{2}{c}{}         &               &   0.48983     &    0.50653     \\
        \multicolumn{1}{l}{O1}  &   x     &     \multicolumn{2}{c}{}         &               &   0.0034      &    0.0472      \\
        \multicolumn{1}{l}{O2}  &   x     &     0.26216     &    0.22212     &               &               &                \\
                                &   y     &     0.76216     &    0.72212     &               &               &                \\
                                &   z     &     0           &    0           &               &   0.9892      &    0.6102      \\
                                                                                                                              \\ \hline
    \end{tabular}
\end{table}

The quality of fit, when modelled using a single phase -- either tetragonal I4/mcm or orthorhombic Ibmm -- is significantly lower as compared to
the two-phase fit. The inset (b1) of figure~\ref{fig:bpbo-txrd-rv}(b) shows the fit of the peaks in the range $76^\circ \leq 2\theta \leq
78^\circ$ when modelled using a single tetragonal phase. A similar fit is observed when the \textit{xrd}{} patterns are refined using a single
orthorhombic phase (not shown). Inset (b2) of figure~\ref{fig:bpbo-txrd-rv}(b) shows the peaks in the range when fit using a diphasic model.  A
sample with tetragonal symmetry shows a single pair of (404) peak (corresponding to Cu K$_{\alpha1}$ and Cu K$_{\alpha2}$)\cite{hashimoto1993}.
On the other hand, in case of orthorhombic symmetry, this peak splits into (404) and (044). To ensure a proper fit, it is necessary to model
data using two phases, which result in the decomposition of twin peaks into its 6 constituent peaks -- one pair corresponding to (404)
reflection from tetragonal phase, one pair each corresponding to (404) and (044) reflections from the orthorhombic phase -- thus indicating the
coexistence of I4/mcm and Ibmm phases.

\begin{figure}
    \centering
    \includegraphics[width=0.80\columnwidth]{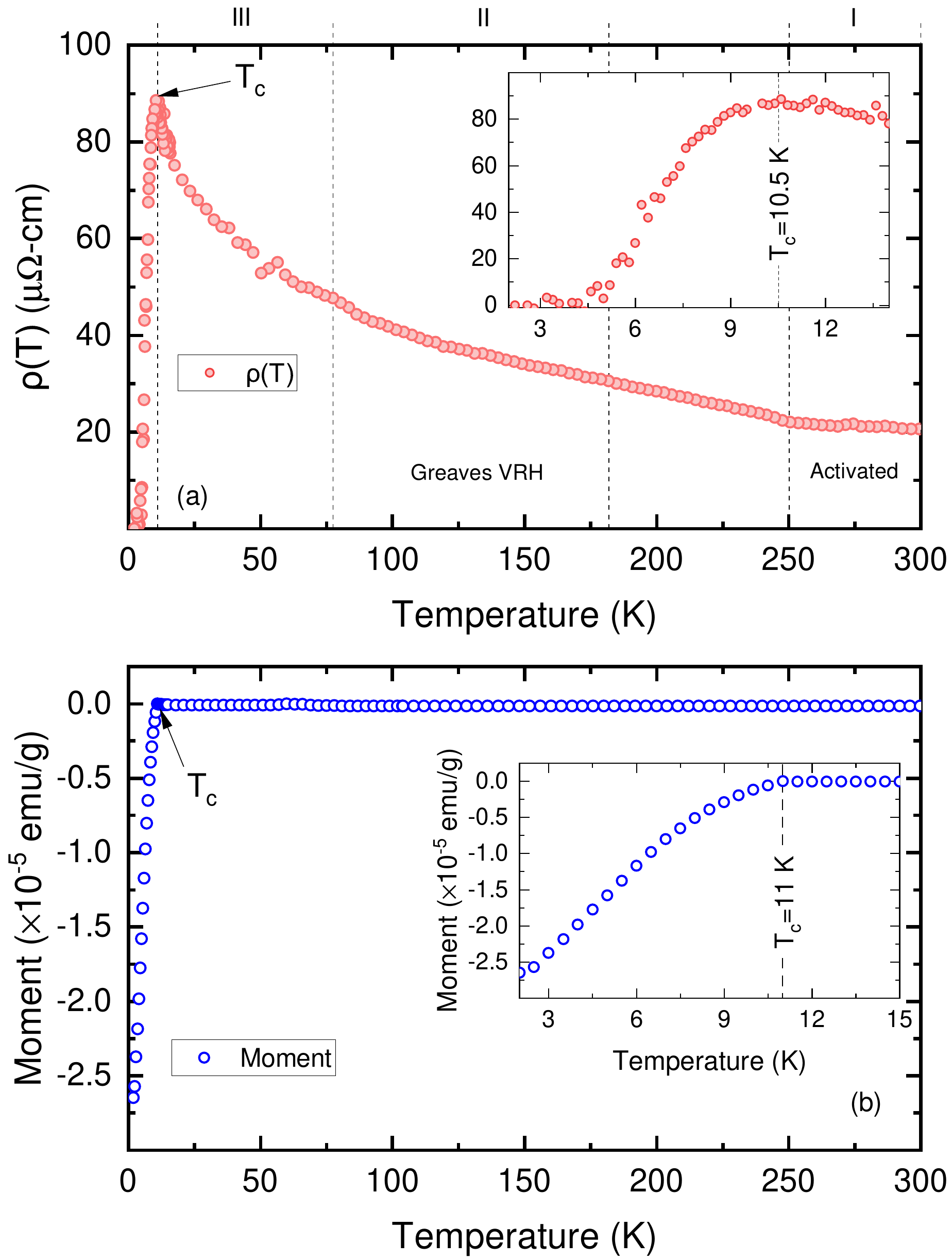}
    \caption{
        (Color online) (a): Variation of resistivity with temperature of BaPb$_{0.75}$Bi$_{0.25}$O$_3$. The inset shows the closer view of
        $\rho(T)$ and $\rho_{300}$ temperature range 2\,K to 15\,K. In both graphs, the open circles represent the experimentally observed data,
        and (b) Variation of magnetic moment with temperature of BaPb$_{0.75}$Bi$_{0.25}$O$_3$. The inset shows the closer view of magnetic
        moment in the temperature range 2\,K to 15\,K.
    }
    \label{fig:bpbo75-rt-mt}
\end{figure}

In figure~\ref{fig:bpbo75-rt-mt}(a), we show the resistivity versus temperature data for the compound under study. As the temperature drops, the
resistivity gradually increases till $\sim$250\,K, below which a change of slope is observed. The resistivity increases rapidly until
$\sim$10.5\,K ($T_C$), below which it drops sharply indicating the onset of superconductivity, as seen in the inset of
figure~\ref{fig:bpbo75-rt-mt}(a). The figure~\ref{fig:bpbo75-rt-mt}(b) shows the field-cooled (FC) magnetic moment of the sample as function of
temperature. As reported in the literature\cite{uchida1988,hashimoto1994}, the sample is diamagnetic in the entire temperature range. The
magnetic moment shows sharp drop at around $\sim$11\,K indicating the onset of the superconducting transition. The combined resistivity and dc
magnetic moment measurements suggest the onset of superconductivity at $T_C\approx 11$\,K.


\begin{figure}
    \centering
    \includegraphics[width=\columnwidth]{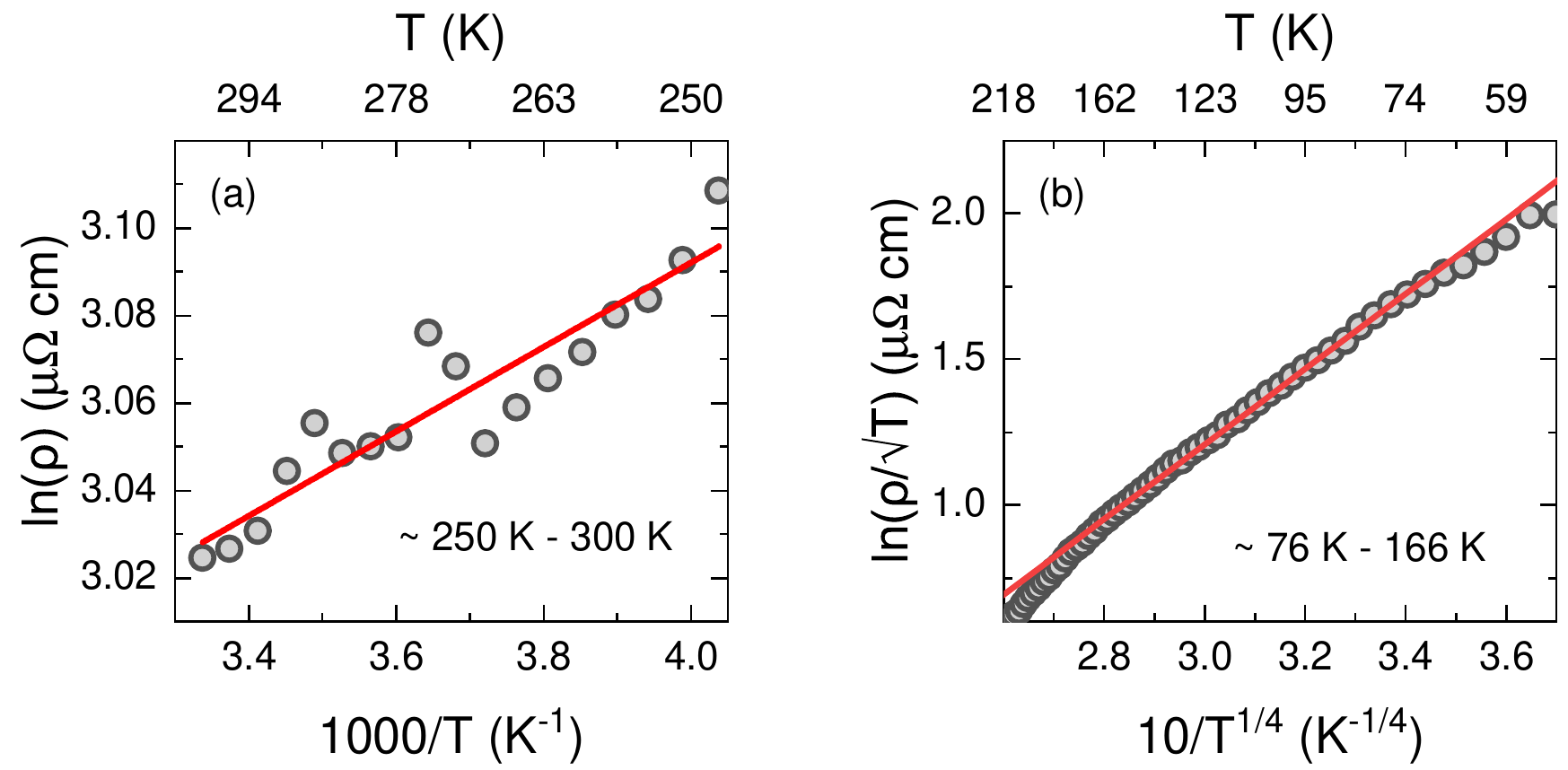}
    \caption{
        (Color online) Variation of resistivity demonstrating
        (a) Arrhenius behaviour in the temperature range 300\,K to 250\,K;
        (b) Greaves' variable-range hopping conduction behaviour in the temperature range below 170\,K to 66\,K.
    }
    \label{fig:bpbo75-rt-activated-vrh}
\end{figure}

Figures~\ref{fig:bpbo75-rt-activated-vrh}(a)-(b) illustrate the fit of the resistivity data in two temperature ranges, using different models.
The resistivity data in figure~\ref{fig:bpbo75-rt-mt}(a) is roughly divided into 3 regions based on this. In region 1, ($\sim250$\,K - 300\,K),
the electrical transport is dominated by activated behaviour, which is appreciable when sufficient thermal energy
is available for the electron to be excited across the band gap\cite{schnakenberg1968, greaves1973, kumara2007}. It is important to note that
the studies by Uchida et al.\cite{uchida1987}, indicate that temperature dependence of normal-state resistivity of BaPb$_{1-x}$Bi$_x$O$_3$ is
like that of semiconductors for $x > 0.20$. Figure~\ref{fig:bpbo75-rt-activated-vrh}(a) shows the variation of $\log\rho$ as a
function of $1/T$, modelled by equation~\ref{eqn:activated}.
\begin{equation}
    \rho(T) = \rho_0 e^{E_G/k_BT}
    \label{eqn:activated}
\end{equation}
where, $\rho_0$ is a pre-exponential factor, $E_G$ is the activation energy, and $k_B = 1.38\times10^{-23}$\,JK$^{-1}$ is the Boltzmann's
constant. From the straight line fit of the data, we obtain an activation energy of $E_g = 7.6$\,meV.

As the temperature drops below 250\,K, it's not possible to model the data using Arrhenius-type behaviour. We observe a transition of the
resistivity behaviour from Arrhenius to variable range hopping (VRH). Mott variable-range hopping model has a characteristic temperature
dependence of resistivity\cite{mott1972, mott2012} given by
\begin{equation}
    \rho(T) =\rho_{0}e^{(T_{M}/T)^{1/(p+1)}}
    \label{eqn:mott-vrh}
\end{equation}
where $p$ is a parameter that describes the dimensionality of the conduction in the system, and $\rho_0$ is a pre-exponential constant. $T_M$ is
the characteristic Mott temperature, which describes the energy barrier of an electron hopping from one localized state to another, given by the
equation~\ref{eqn:tm}.
\begin{equation}
    T_M = \frac{18.1}{k_B\xi^3 N(E_f)}
    \label{eqn:tm}
\end{equation}
Here, it is important to note that it was not possible to fit the entire temperature range below 250\,K, till the $T_C$, with a single curve. It
was necessary to split them into separate regions with distinct characteristic temperatures and localization lengths. However, while we were
able to fit the data using equation~\ref{eqn:mott-vrh} mathematically, we found that the ratio of mean hopping distance to electron localization
length, given by equation~\ref{eqn:local_length}, is much less than 1, which is an indication that equation~\ref{eqn:mott-vrh} is not
applicable in our case\cite{rosenbaum1991}.
\begin{equation}
    R_M/\xi = \frac{3}{8}\left(\frac{T_M}{T}\right)^{1/4}
    \label{eqn:local_length}
\end{equation}

In the intermediate temperature range, the resistivity data can be fit using Greaves VRH\cite{greaves1973}. The acoustic phonon contribution to
the resistivity\cite{schnakenberg1968, greaves1973}, takes the form as given in equation~\ref{eqn:rho},
\begin{equation}
    \rho(T) = AT^{1/2}e^{(T_G/T)^{1/4}}
    \label{eqn:rho}
\end{equation}
where, $A$ is the pre-exponential constant and $T_G$ is the characteristic temperature of Greaves VRH.

For Greaves VRH to be valid, the plot of $\ln(\rho/\sqrt{T})$ versus $T^{-1/4}$ should be linear. In this intermediate temperature range
($\Theta_D/4 < T < \Theta_D/2$) where the deviation from the fit is observed has been taken to be $\Theta_D/2$ in the higher temperature limit,
and $\Theta_D/4$ in the lower temperature limit\cite{sakata1999, kumara2007}. Below this temperature, we have been unable to model the
resistivity using this equation. Thus, the best fit was observed in the temperature range $\sim$76\,K to 166\,K, and we obtained the
characteristic temperature of $T_G = 27563$\,K. Using the value of $N\left(E_F\right) = 2.2\times10^{21}$ states/cm$^3$, as reported by
Kitazawa, et al.\cite{kitazawa1985}, we obtain the electron localization length of $\xi = 15.49 $\,\textup{\AA}. The deviation of the
resistivity from the fit begins at $\sim$76\,K at the lower end, and $\sim$166\,K at the higher end. Thus, we may estimate that the Debye
temperature of the compound lies between $\sim$304\,K to 332\,K.

\subsection{\label{subsec:txrd}
Temperature dependent \textit{xrd}{} studies} 
\begin{figure*}
    \centering
    \includegraphics[width=0.65\textwidth]{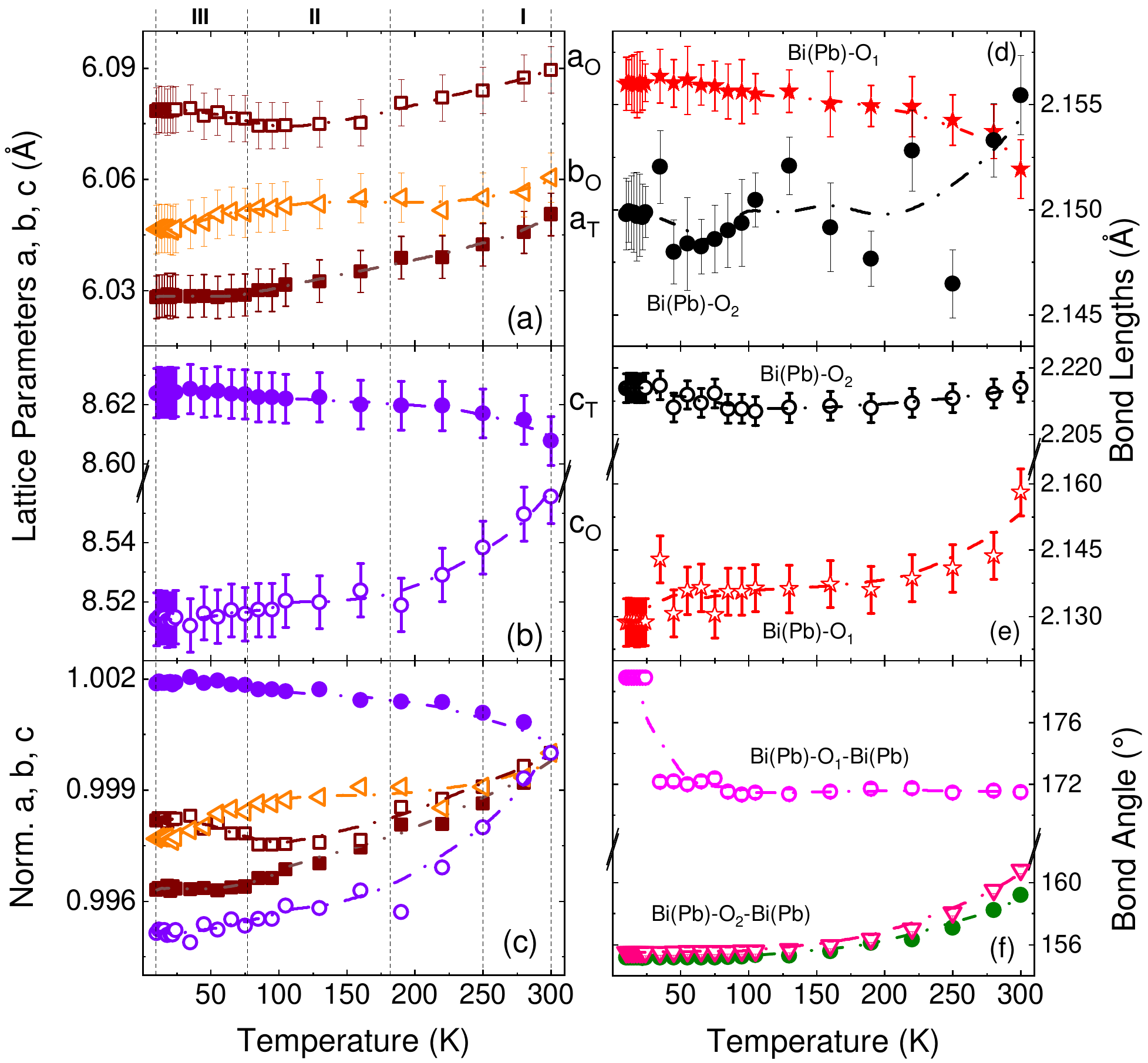}
    \caption{
        (Color online) Variation of lattice parameters with temperature for tetragonal phase (closed symbols) and orthorhombic phase (open
        symbols) of BaPb$_{0.75}$Bi$_{0.25}$O$_3$.
        (a) and (b) show the evolution of $a$ (square), $b$ (triangle) and $c$ (circle) lattice parameters, respectively.
        (c) shows the temperature variation of the lattice parameters normalized with respect to their respective values at RT.
        (d) shows the variation in bond lengths of Bi/Pb and the apical (star) and basal (circle) oxygens in the I4/mcm phase.
        (e) shows the variation in bond lengths of Bi/Pb and the apical (star) and basal (circle) oxygens in the Ibmm phase.
        (f) shows the variation of the angle between two Bi(Pb) and apical and basal oxygens.
        Dashed lines are a visual guide indicating the overall behaviour of the various parameters.
    }
    \label{fig:bpbo75-txrd-params-all}
\end{figure*}

\begin{figure}
    \centering
    \includegraphics[width=\columnwidth]{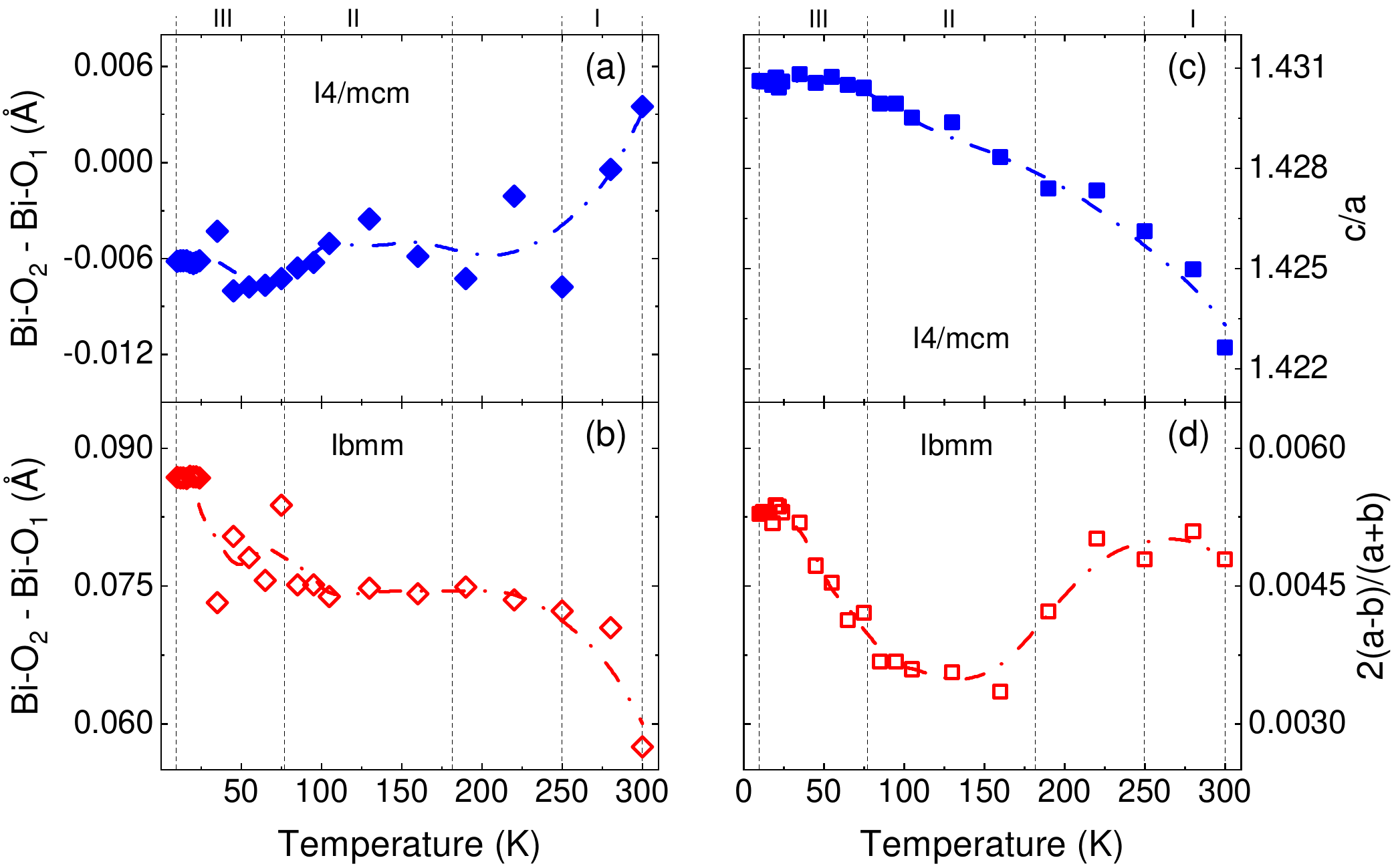}
    \caption{
        (Color online) Temperature variation of
        (a) basal and apical bond length difference in tetragonal phase,
        (b) basal and apical bond length difference in orthorhombic phase,
        (c) tetragonal strain , $c/a$ and
        (d) orthorhombic strain, $2(a-b)/(a+b)$.
        Dashed lines are a visual guide indicating the overall behaviour of the various parameters.
    }
    \label{fig:bpbo75-txrd-strain}
\end{figure}

\begin{figure}
    \includegraphics[width=\columnwidth]{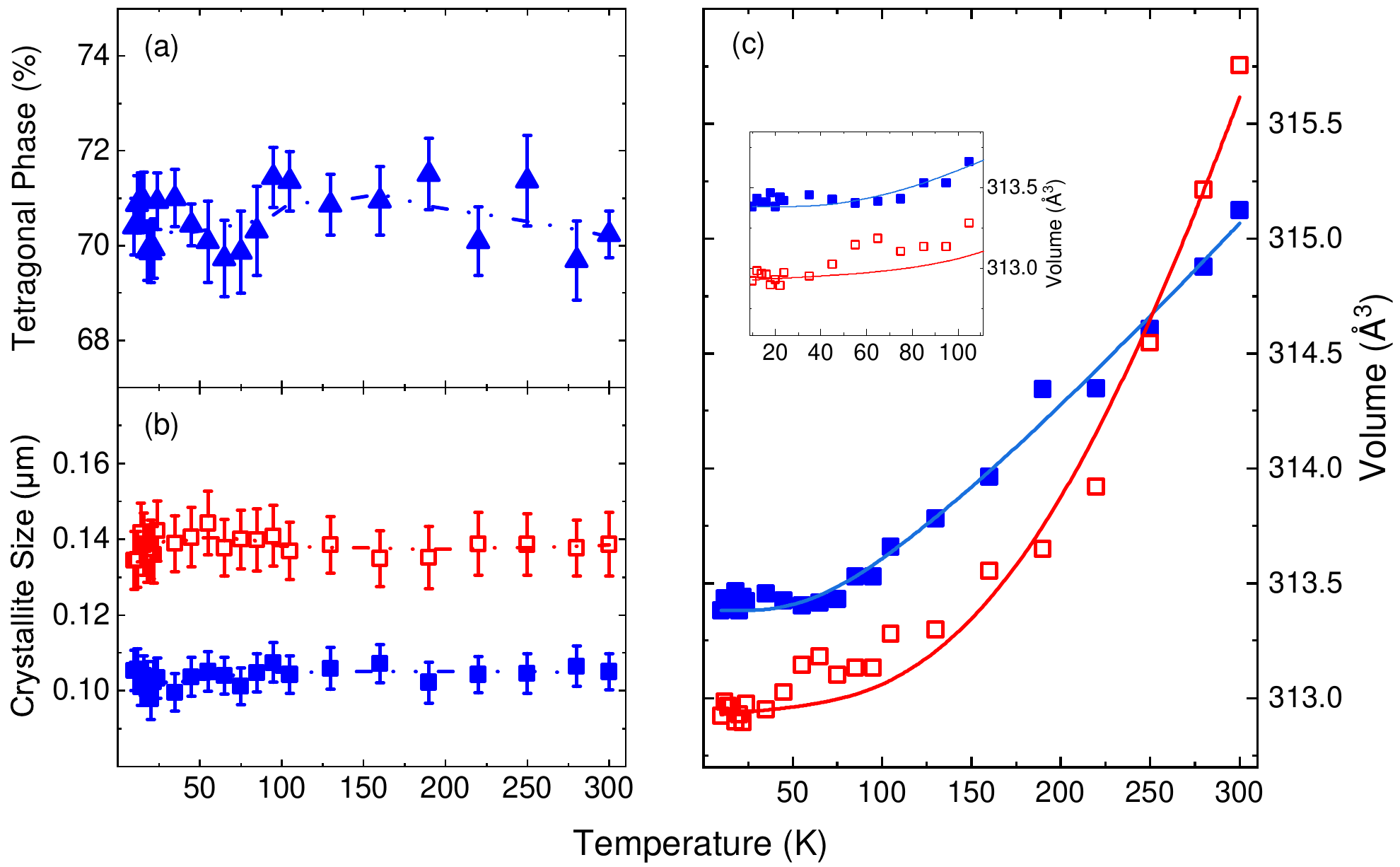}
    \caption{
        (Color online) The temperature variation of
        (a) tetragonal phase faction
        (b) crystallite size of tetragonal (closed squares) and orthorhombic (open squares) phases calculated from the Scherrer equation.
        (c) volumes of tetragonal and orthorhombic phases (closed and open square, respectively), and the Debye fit to the anharmonic part of
        the lattice contribution to the unit cell volumes (solid lines). The inset shows the close-up of the volumes in the temperature
        range 10\,K to 110\,K.
    }
    \label{fig:bpbo75-tetra-fraction}
\end{figure}

To understand the structural link with the transport properties and understand the behaviour of
phase separation, we have carried out temperature dependent \textit{xrd} on the compound.
Figures~\ref{fig:bpbo75-txrd-params-all}-\ref{fig:bpbo75-tetra-fraction} show the variation of the lattice parameters, bond lengths, bond
angles, lattice strain, and unit cell volume with temperature. The \textit{xrd}{} patterns were analysed using Rietveld profile refinement and
the structural parameters were obtained as a function of temperature. The goodness of fit obtained is in the range 1.45 -- 1.63 for all
temperatures. As the temperature drops from 300\,K to 10\,K, we observe that the number of peaks remain the same. This suggests the absence of
any structural transitions at lower temperatures, in contrast to some of the earlier studies\cite{oda1985,asano1988}. Moreover, we observe that
the fraction of the tetragonal phase is $\sim$70\%, and is largely temperature independent (see figure~\ref{fig:bpbo75-tetra-fraction}(a)).

Keeping these results in mind, revisiting the resistivity studies of the previous section gives an opportunity to understand the high
transition temperature of VRH from Arrhenius type behaviour. Firstly, at low temperature, the thermal energy available for excitation of
electron is insufficient, thus leading the electrical transport behaviour away from activated behaviour. In such a scenario, it is favorable
for the electron to hop to a site farther than the nearest neighbour with a lower potential, giving rise to observed VRH
behaviour\cite{kumara2007}. Furthermore, we hypothesize that this high transition temperature is due to the role of disorder in the system. Two
leading contributions to disorder are (a) the structural phase coexistence of orthorhombic and tetragonal phases at all temperatures, with
fraction of the tetragonal phase constant at $\sim$70\%, and largely temperature independent, and (b) the composition lying within the
transition region between metal and semiconducting regions. The work on BaPb$_{1-x}$Bi$_{x}$O$_3${} by Uchida et al.\cite{uchida1987} suggests
that the compound under study lies in the region where both metal and semiconducting regions co-exist.

Figures~\ref{fig:bpbo75-txrd-params-all}(a)~and~\ref{fig:bpbo75-txrd-params-all}(b) illustrate the variation of the lattice parameters of I4/mcm
and Ibmm phases with temperature. In order to facilitate the comparison of the structural data with resistivity measurements, the entire
temperature range have been divided into three regions, identical to that in figure~\ref{fig:bpbo75-rt-mt}(a).
In the activated region, the $a$ parameter of the tetragonal phase, decreases by 0.0014\%, whereas $c$ increases by 0.001\%. As the temperature
drops to region II, the decrement in $a$ is 0.0022\%, and the increment in $c$ is 0.001\%. In region III, the lattice parameters are nearly
constant within the range of experimental error.

The antiphase Bi(Pb)-O$_6${} octahedral rotations about the $c$-axis increase with decreasing temperature. The decreasing Bi(Pb)-O$_2$-Bi(Pb)
bond angle results in the contraction of the $a$ parameter. This reduction in the bond angle effectively contracts the interstitial volume
around the Barium atom, thus increasing the O$_1$-Ba-O$_2$ angle, which drives the increase in the $c$ parameter with decreasing temperature.

The lattice parameters of the Ibmm phase, however, paint a more complex picture. The figure~\ref{fig:bpbo75-txrd-params-all}(a)~and~(b)
respectively show the variation of $a$, $b$, and $c$ parameters of the Ibmm phase. The lattice parameters $a$ and $c$ reduce monotonically until
$\sim$75\,K, dictated by the reducing bond lengths and bond angles; further, $b$ decreases monotonically until the beginning of region II, and
remains constant in region II. Below $\sim$75\,K, $a$ increases, while $b$ decreases; $c$ remains nearly constant. The interplay between the
bond lengths and bond angles leading to the observed behaviour of the lattice parameters are easily understood from graphs in panels (e) and (f)
of figure~\ref{fig:bpbo75-txrd-params-all}.

From figures~\ref{fig:bpbo75-txrd-strain}(a)~and~(b), we clearly observe that the nature of Bi(Pb)-O$_6${} octahedra is different in the two
phases. In the case of tetragonal phase, there are two long apical Bi-O bonds and four short basal Bi-O bonds, while in the case of orthorhombic
phase, the trend is reversed.  It is interesting to note that, Gallo et al.\cite{giraldo2015}, have observed that the phase separation takes the
form of partially disordered nanoscale stripes. It is possible that the difference in the nature of distortion could be one of the reasons for
such stripes. These experiments were carried out by the authors on single crystalline BaPb$_{0.75}$Bi$_{0.25}$O$_3$ and have observed nanoscale
structural phase separation.

To understand the nature of phase separation in the present work, one can propose two scenarios\cite{chmaissem2003, sharma2022} (a) two phases
coexisting in the same grain; (b) two phases lying in two different grains. In the first scenario, it is expected that one phase grows at the
cost of the other while in the second case the phase fraction remains the same. Towards this, we have calculated the average size of the
structural phases as a function of temperature using Scherrer formula\cite{patterson1939} (equation~\ref{eqn:scherrer}).
\begin{equation}
    B=\frac{K\lambda}{\beta\cos\theta}
    \label{eqn:scherrer}
\end{equation}
where, $K=0.9$ is a constant, $\lambda = 1.5405$\,\textup{\AA}{} is the x-ray wavelength, $\beta$ is the FWHM of the peak at angle $2\theta$.
Our results show that the size of the structural phases remains the same through temperature range of study. This behaviour suggests the second
scenario proposed above. However, temperature dependent transmission electron microscopy (TEM) studies will be helpful to unravel the nature of
phase separation in this polycrystalline compound.

Graphs in figure~\ref{fig:bpbo75-txrd-strain}(c)~and~(d) show the temperature variation of tetragonal and orthorhombic strain. In the forth
coming section, we will be using the term lattice strain synonymous to tetragonal/orthorhombic strain. In the tetragonal phase
(figure~\ref{fig:bpbo75-txrd-strain}(c)), we observe the lattice strain increases with decrease in temperature and in the orthorhombic phase
(figure~\ref{fig:bpbo75-txrd-strain}(d)), the lattice strain decreases until around 125 K and below this temperature it is found to increase. It
is interesting to note that in the temperature range close to $T_C$, the lattice strain is maximum possibly playing a role as precursor for
superconductivity. Such behaviour has been observed in the case of YBa$_2$Cu$_3$O$_7$ compound\cite{horn1987}, where the orthorhombic strain is
found to be maximum around $T_C$ with little to no anomaly in the unit cell volume. However, in the present study we observe an anomaly in the
unit cell volume around 35 K and 65 K, in the tetragonal and orthorhombic phases, respectively, (see figure~\ref{fig:bpbo75-tetra-fraction}(c)).
In this figure, the anharmonic part of the lattice contribution to the unit cell volumes was obtained by fitting with Debye model, as given
by equation~\ref{eqn:debye-fit}.
\begin{equation}
    V \approx V_0 +\frac{9\gamma Nk_B}{B} T \left(\frac{T}{\Theta_D}\right)^3 \int_{0}^{\Theta_D/T}{\frac{x^3}{e^x-1}\,dx}
    \label{eqn:debye-fit}
\end{equation}
where, $V_0$ is the volume at absolute zero, $N$ is the number of atoms per unit cell, $\gamma$ is the Gr\"unesian parameter and $\Theta_D$ is the
Debye temperature. The three fitting parameters -- $V_0$, $\left(9\gamma Nk_B\right)/B$ and $\Theta_D$ -- were determined using the least square
fitting, are shown in table~\ref{tbl:debye-fit}. We readily see that the $\Theta_D^{\rm tetra} = 330.4$\,K, the Debye temperature of the
majority phase fraction of the compound, obtained from the fit matches well with the estimate obtained from the resistivity data, as previously
discussed.
\begin{table}
    \caption{Parameters of Debye fit obtained from fitting of tetragonal and orthorhombic phase volumes.}
    \label{tbl:debye-fit}

    \begin{tabular}{lccccccc}\hline
        Phase         & $\quad$  &  $V_0$   $\quad$  &   $\left(9\gamma Nk_B\right)/B\quad$ &   $\Theta_D$    &   $\quad$     \\   \hline
        Tetragonal    & $\quad$  &  313.38  $\quad$  &   0.026       $\quad$                &   330.4         &   $\quad$     \\
        Orthorhombic  & $\quad$  &  312.92  $\quad$  &   0.098       $\quad$                &   916.8         &   $\quad$     \\   \hline
    \end{tabular}
\end{table}


\section{\label{sec:txps}Electronic structure studies}

\begin{figure*}
    \centering
    \includegraphics[width=\textwidth]{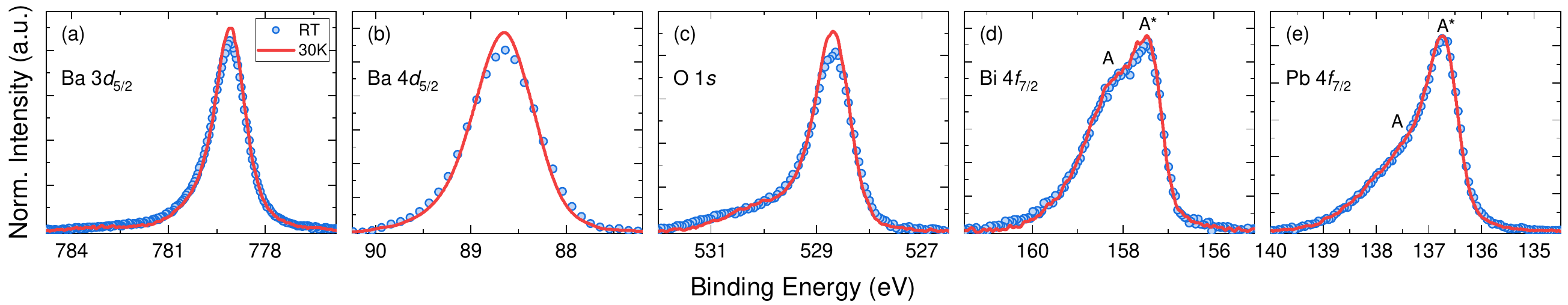}
    \caption{ Core level spectra of BaPb$_{0.75}$Bi$_{0.25}$O$_3$ at 300\,K (open circles) and 30\,K (solid lines). (a) Ba $3d_{5/2}$ (b) Ba
        $4d_{5/2}$ (c) O $1s$ (d) Bi $4f_{7/2}$ and (e) Pb $4f_{7/2}$. The labels $A$ and $A^*$ in panels (d) and (e) indicate the positions of
        poorly-screened and well-screened peaks, respectively. }
    \label{fig:bpbo-txps}
\end{figure*}

The results discussed thus far indicate that the disorder plays a very important role in the sample, especially in the region just above $T_C$.
Signature of disorder is visible not only in transport measurements, but also is expected to manifest in electronic structure. According to BCS
theory, good conductors do not show superconductivity because of weak electron phonon coupling. In the present compound, the observed
resistivity values lie in the range of metals. It will be interesting to study the temperature dependent behaviour of the electronic states
close to the Fermi level and also behaviour of core levels prior to $T_C$. To visualize and understand the electronic structure of
BaPb$_{1-x}$Bi$_{x}$O$_3$, we have performed photoemission spectroscopy measurements on our sample at various temperatures between 300\,K and
30\,K, using Al K$_\alpha$ x-ray radiation, He$_{\mathrm I}${} and He$_{\rm II}${} ultraviolet radiations. Figure~\ref{fig:bpbo-txps} shows
the core levels of the compound collected at RT and 30\,K.

\subsection{Core level studies}
Panels (a)–(e) of figure~\ref{fig:bpbo-txps} show the core level spectra of Ba 3$d_{5/2}$, Ba 4$d_{5/2}$, O 1$s$, Bi 4$f_{7/2}$, and Pb
4$f_{7/2}$, respectively, at 300\,K and 30\,K. Our results show that with decrease in temperature there is a decrement in the width of the core
level peaks. This is very clearly observed in the case of Ba and O 1s core levels. The Bi and Pb 4$f_{7/2}$ core levels exhibits two features
labelled as A and A*. Such features have been observed in Bi core level spectra by Prachi et al.\cite{telang2022} in the case of Bi based
pyrochlore iridate where the sample is metallic and have attributed the features A and A* to be of poorly and well screened features. Such two
features have also been observed in the case of Pb 4$f_{7/2}$ by Payne et al.\cite{payne2007}, in the case of PbO$_{2}$ sample that is metallic.
It is interesting to note that in the case of BaBiO$_{3}$, that is semi conducting, only poorly screened feature is observed\cite{bharath2019,
plumb2016}.  The well screened feature arises when the Bi/Pb 4$f_{7/2}$ core hole is screened by the transfer of electrons from the ligand and
the poorly screened feature arises when no such transfer occurs.

\subsection{\label{subsubsec:tpes-vb}xps Valence Band and Band Strucutre Studies}

\begin{figure}
    \centering
    \includegraphics[width=\columnwidth]{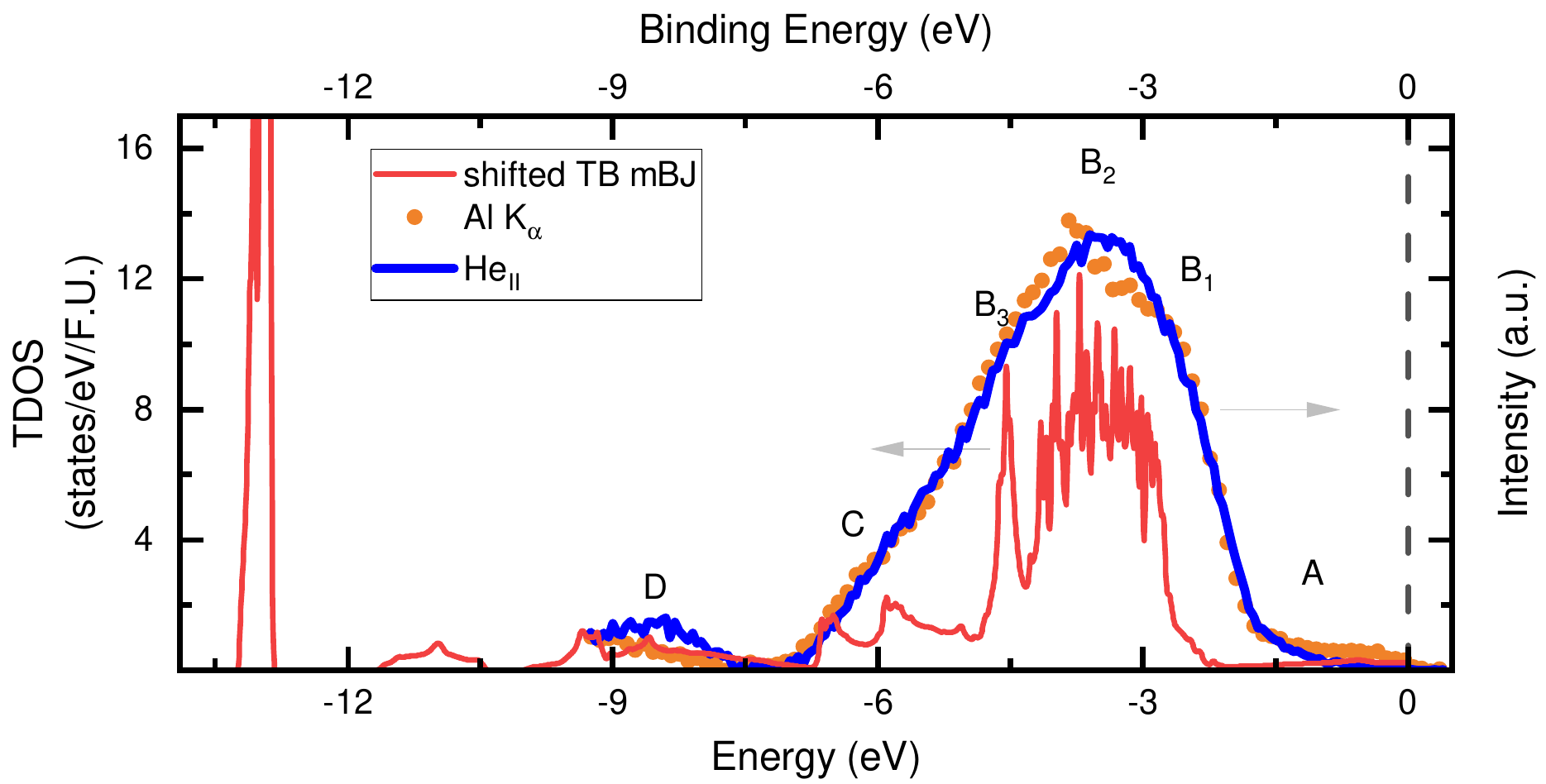}
    \caption{
        (Color online) A comparison of Al K$_\alpha$ spectrum (closed circles, orange), He$_{\rm II}${} spectrum (solid line, blue) and the
        total DOS (solid line, red) obtained from the TB-mBJ calculations performed on tetragonal phase of BaPb$_{0.75}$Bi$_{0.25}$O$_3$.
        The total DOS is shifted by -0.8 eV to match the experimental Al K$_\alpha$ spectrum. The inset shows the finite DOS at Fermi level
        using Al K$_\alpha$ radiation.
    }
    \label{fig:expt-dft}
\end{figure}

\begin{figure}
    \centering
    \includegraphics[width=\columnwidth]{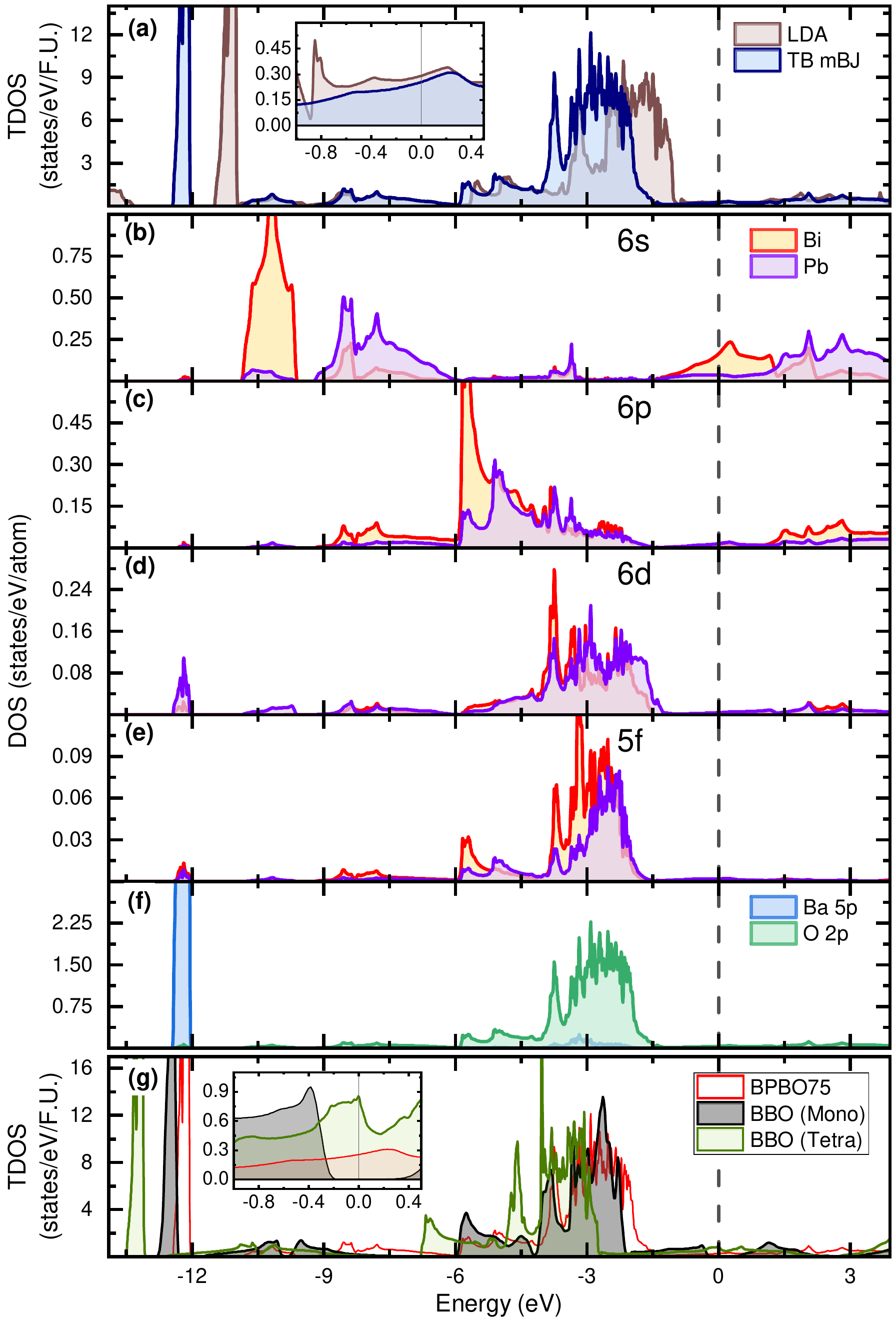}
    \caption{(Color online) DFT calculations performed on tetragonal phase of BaPb$_{0.75}$Bi$_{0.25}$O$_3$. Panel
        (a) shows the comparison between LDA (brown) and TB-mBJ (blue) calculations; inset shows the region near $E_F$, where finite DOS is
        clearly visible;
        panels (b)--(e) show the partial DOS (PDOS) of Bi (red) and Pb (violet) $6s$, $6p$, $6d$ and $5f$ states, respectively obtained using
        TB-mBJ calculations;
        panel (f) shows the PDOS of Ba $5p$ (light blue) and O $2p$ (green) states obtained using TB-mBJ calculations .
        Panel (g) shows the TDOS obtained from the TB-mBJ calculations for BaBiO$_3$ in the monoclinic phase (black),
        BaPb$_{0.75}$Bi$_{0.25}$O$_3$ in the tetragonal phase (red), and BaBiO$_3$ using the structural parameters of
        BaPb$_{0.75}$Bi$_{0.25}$O$_3$ (green).
    }
    \label{fig:dft-bpbo75}
\end{figure}

The comparison of xps and He$_{\rm II}${} valence band spectra collected at 300\,K is shown in figure~\ref{fig:expt-dft}. Finite DOS at $E_F$ is
clearly visible in both xps and He$_{\rm II}${} spectra. We observe six features in the spectra labelled as A, B$_1$, B$_2$, B$_3$, C and D. The
xps valence band represents the bulk features of the sample. To identify the features, band structure calculations were carried out on
BaPb$_{0.75}$Bi$_{0.25}$O$_3$, in the tetragonal phase using DFT method, as shown in figure~\ref{fig:dft-bpbo75}. Our results show that the
total density of states (TDOS) of TB-mBJ is shifted towards lower energy as compared to the LDA calculations, as seen in panel (a) of
figure~\ref{fig:dft-bpbo75}. The partial density of states (PDOS) are plotted in panels (b)--(f) of figure~\ref{fig:dft-bpbo75}. While LDA
calculations generally sufficient in case of metallic compounds, it has been well documented that LDA underestimates the band gap at the fermi
level\cite{perdew1983, sham1983} of semiconducting or insulating compounds. This was the case of the parent compound, BaBiO$_3$, where the gap
obtained in the case of LDA calculations was underestimated as compared to the TB-mBJ calculations\cite{bharath2019}. However, the compound
under study, BaPb$_{0.75}$Bi$_{0.25}$O$_3$, exhibits both metallic and semiconducting behaviours, as discussed previously. Further, we're also
probing the effects of Pb doping in BaBiO$_3$ to understand how the band gap observed in BaBiO$_3$ closes upon Pb doing. Thus, TB-mBJ
calculations have been employed to understand the electronic structure of BaPb$_{0.75}$Bi$_{0.25}$O$_3$. The experimental spectra were matched
with the TB-mBJ calculations with a finite energy shift. Towards this, in the present work, a rigid shift of -0.8 eV was given to the DOS
obtained from the TB-mBJ calculation to match with the experimental spectrum.

Our results show that there is finite DOS at the Fermi level that is in line with the experimental spectra. The calculated DOS can be divided
into five regions. In region I (-14 to -12 eV), there is significant contribution from Ba $5p$ states and weak contributions from Bi/Pb $6s$ and
Pb $6d$ states. Region II (-12 to -7 eV) covers dominant contributions from Bi/Pb $6s$ and weak contributions from rest of the DOS. Region III
(-7eV to -1.5 eV) has contributions from Bi/Pb $6s$, $6p$, $6d$, $5f$ and O $2p$ states. In region IV (-1.5 to $E_F$) there is dominant
contribution from Bi $6s$ and weak contributions from other states. The region V (above $E_F$ to 2\,eV) has dominant contribution from Pb
$6s$ and weak contributions from rest of the states. We now identify the different features based on the above results. The feature $D$
represents surface cleanliness. In the case of xps, the intensity of this feature is low as compared to the He$_{\rm II}${} spectra due to
higher sensitivity to surface oxygen. The features B$_1$, B$_2$ and B$_3$ covers region III and the feature $A$ encompasses region II. The
photoionisation cross section for Al K$_\alpha$ source is more for Bi/Pb $6s$ as compared to the photoionisation cross section of O $2p$, while
in the case of He$_{\rm II}${} spectra, the reverse is true. Hence, the higher intensity of feature $A$ observed in xps as compared to He$_{\rm
II}${} spectra suggests the contributions from Bi/Pb $6s$ states.

Pb doping introduces a significant change in the electronic structure near $E_F$ as compared to BaBiO$_3${}. An insulating gap at $E_F$ observed
in BaBiO$_3${}\cite{bharath2019} closes on 75\% Pb doping. The region between -3 to -1.4 eV which was identified as feature $B$ in
BaBiO$_3${}\cite{bharath2019} splits into 3 features upon Pb doping, namely, B$_1$, B$_2$ and B$_3$. All the three features have dominant
contributions from O $2p$ states. In figure~\ref{fig:dft-bpbo75}\,(g), we have presented the TB-mBJ calculations of monoclinic BaBiO$_3$
(labelled as BBO (Mono)), BaBiO$_3${} using the structural parameters of BaPb$_{0.75}$Bi$_{0.25}$O$_3$ (labelled BBO (Tetra)) and
BaPb$_{0.75}$Bi$_{0.25}$O$_3$. By performing TB-mBJ calculations on BaBiO$_3$ using the structural parameters of BaPb$_{0.75}$Bi$_{0.25}$O$_3$,
we intend to understand the origin of states at $E_F$. Our results show that there is finite DOS at $E_F$ in this case. This suggests that the
finite DOS appearing at $E_F$ is mainly driven by the crystal structure. Further, the DOS at Fermi level was observed to decrease upon Pb
doping, as seen clearly in the inset of figure~\ref{fig:dft-bpbo75}\,(g). To ascertain that the position of bismuth does not play an important
role in the electronic structure of BaPb$_{0.75}$Bi$_{0.25}$O$_3$, we repeated the TB-mBJ calculations by substituting the bismuth atom in each
of the different positions of lead. The total DOS was observed to be identical in all the cases.

\subsection{\label{subsubsec:tups-vb}UPS Valence Band Studies near $E_F$}

\begin{figure}[ht]
    \centering
    \includegraphics[width=\columnwidth]{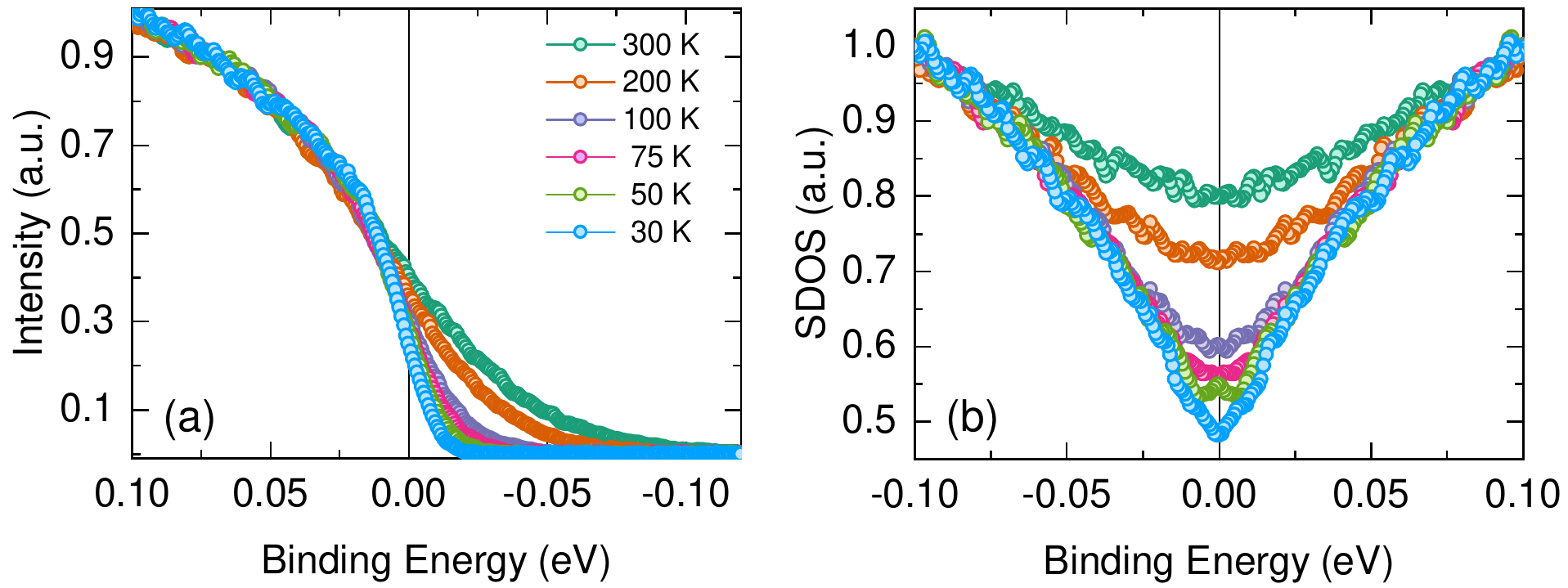}
    \caption{ (a) He$_{\mathrm I}${} Spectra of BaPb$_{0.75}$Bi$_{0.25}$O$_3$ collected at various temperatures in a narrow range; fermi level,
        $E_F$ is at binding energy $ = 0$\,eV (b) Spectral DOS obtained from the symmetrization of the He$_{\mathrm I}${} spectra at different
        temperatures, shown in panel (a).}
    \label{fig:bpbo75-ups-vbsymm}
\end{figure}

\begin{figure}[ht]
    \centering
    \includegraphics[width=\columnwidth]{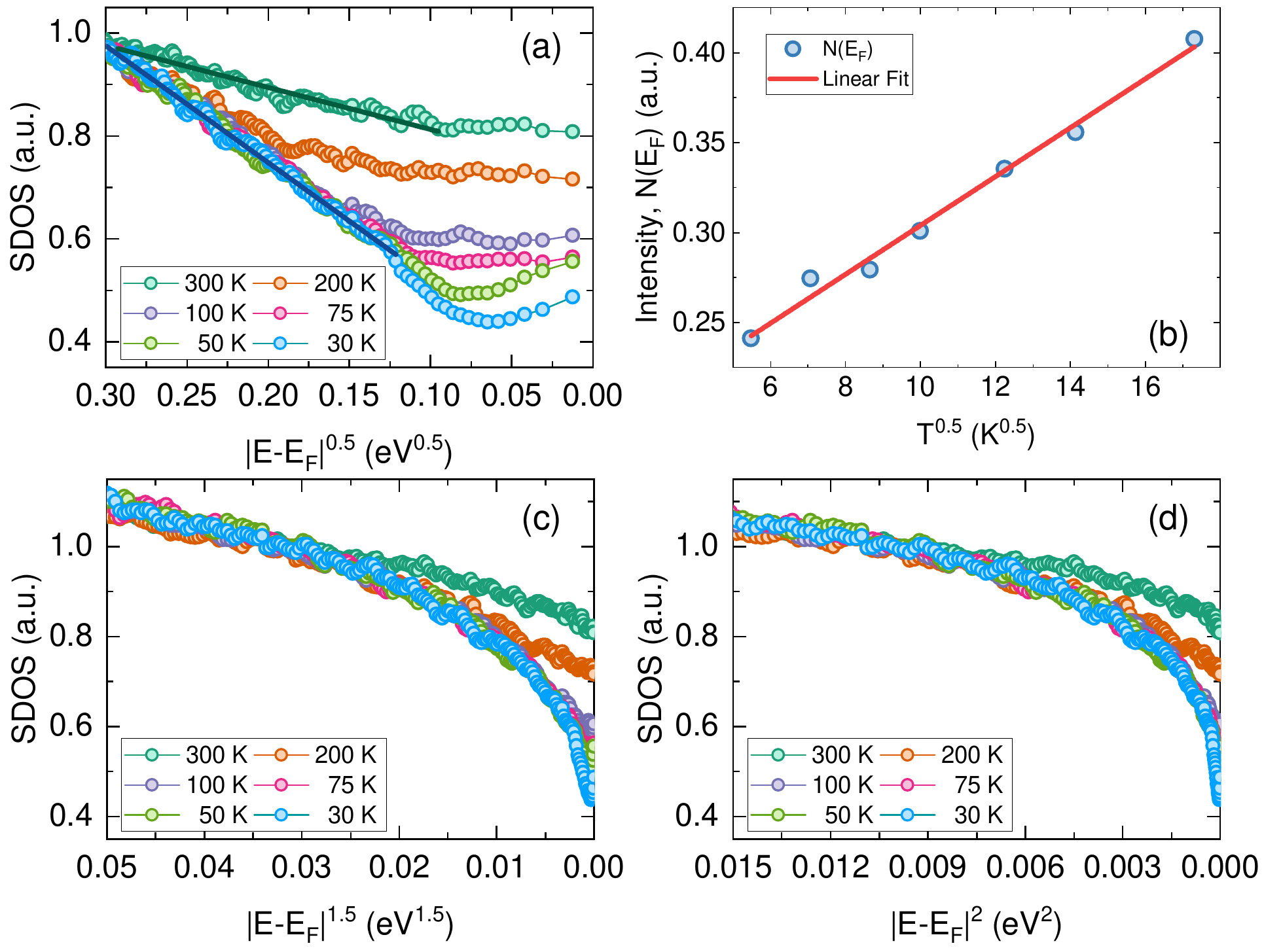}
    \caption{ (a) Spectral DOS (open circles) as a function of $\left|E-E_F\right|^{1/2}$, (b) Intensity at $E_F$. $N(E_F)$, as a function of
        $T^{1/2}$ (open circles). The linear fit (solid lines) to SDOS in panel (a) and $N(E_F)$ in panel (b) indicate that the sample is a
        disordered metal. (c) SDOS (open circles) as a function of $\left|E-E_F\right|^{3/2}$, and (d) SDOS (open circles) as a function of
        $\left|E-E_F\right|^{2}$, }
    \label{fig:bpbo75-ups-vb-disorder}
\end{figure}

Having identified the states that contribute to the Fermi level ($E_F$), we now look into the evolution of the states close to $E_F$. In
figure~\ref{fig:bpbo75-ups-vbsymm}(a), we show the temperature dependent valence band collected using HeI spectra. All the spectra are
normalised at 100\,meV. Our results show that all the spectra show a finite intensity at $E_F$ and this intensity is found to decrease on
lowering the temperature. To observe more clearly the temperature behaviour of the states close to $E_F$, the spectral density of states was
obtained by dividing the experimental spectra by the Fermi Dirac distribution and symmetrising it, figure~\ref{fig:bpbo75-ups-vbsymm}(b). The
figure shows significant reduction in the intensity as a dip in the vicinity of $E_F$ with decrease in temperature. This dip increases as the
temperature drops and a systematic increase in the spectral DOS above 100 meV is observed with decrease in temperature suggesting the
stabilization of pseudogap.

It is important to note that Matsuyama et al.\cite{matsuyama1989} have collected He I and II spectra on BaPb$_{0.85}$Bi$_{0.15}$O$_3$ at around
300\,K and 180\,K. Based on the Fermi edge observed in this compound, they proposed that the superconductivity would be driven by cooper
pairing of electrons in Fermi liquid states. To check for this possibility, the SDOS was plotted as a function of $(E-E_F)^n$. Our results show
that the value of $n$ thus obtained was 0.5 suggesting the role of disorder existing in the compound, as seen in
figure~\ref{fig:bpbo75-ups-vb-disorder}(a). Earlier studies\cite{{sarma1998,kobayashi2007}} have shown that such decrement in the intensity
could arise due to the localization of the states at $E_F$ induced by disorder. Under this situation, it is expected that the DOS at $E_F$
follows $N(E_F) = a + b\sqrt{T}$, where $T$ is the temperature; such a behaviour can be clearly observed in panel (b) of
figure~\ref{fig:bpbo75-ups-vb-disorder}, thus suggesting that the compound is a disordered metal for $T>T_C$. However, to observe whether there
is any characteristic peak observed close to $E_F$ representing the superconducting state, experiments below $T_C$ are required.


\section{Summary}
In conclusion, we have studied the structural, transport and electronic properties of the polycrystalline superconducting
BaPb$_{0.75}$Bi$_{0.25}$O$_3${} sample. The analysis of the temperature dependent resistivity data reveals the compound undergoes a
superconducting transition around 11\,K. The VRH behaviour of the resistivity at low temperatures in an indication that disorder plays a very
important role in the electronic properties of the compound. We believe that the structural dimorphism present in the sample at all
temperatures, along with the coexistence of metallic and semiconducting regions in the sample are the likely causes for disorder in the
compound under study. To explore the effect of phase separation, structural studies were conducted using \textit{xrd}{} which reveal that the
compound is dimorphic at all temperatures ranging from 300\,K down to 10\,K. Further, the evolution of the tetragonal and orthorhombic strain
with temperature indicate that the strain is maximum as we approach the $T_C$. The observation of well screened features in the Bi and Pb spin
orbit split 4f core level suggests the metallic nature of the sample. Additionally, we observed a finite density of states at Fermi level. Our
band structure studies suggest that the closing of the gap upon Pb doping is due to structural transition from the monoclinic I2/m phase to the
tetragonal I4/mcm phase. The temperature-dependent behaviour of the electronic states close to the fermi level was studied using UPS
measurements. The signature of disorder was observed in the form of $|E-E_F|^{1/2}$ dependence of SDOS near the Fermi level suggests the
compound under study for $T>T_C$ is a disordered metal. Furthermore, the DOS at Fermi level decreased progressively with temperature -- an
evidence of opening of disorder-induced pseudogap. Our combined crystal structure, transport and electronic structure studies suggest that
lattice strain and disorder act as precursor to superconductivity in BaPb$_{0.75}$Bi$_{0.25}$O$_3$ compound.

\section{Acknowledgements}
The authors, Bharath M and R Bindu thank Science and Engineering Research Board (SERB), Department of Science and Technology, Government of
India for funding this work. This work is funded under the SERB project sanction order No. EMR-2016-001144.

\bibliographystyle{apsrev}
\bibliography{Third}

\end{document}